\documentstyle[psfig,sprocl]{article}

\def\be{\begin{equation}}
\def\ee{\end{equation}}
\def\bea{\begin{eqnarray}}
\def\eea{\end{eqnarray}}
\begin{document}

\noindent {\footnotesize 
Ed.: L. Matsson, 
``Nonlinear Cooperative Phenomena in Biological Systems'', \\[-1mm]
Proc. of the Adriatico Research Conference, 
ICTP, Trieste, Italy, 19-22 August 1997 \\[-1mm]
(World Scientific, Singapore, 1998) pp. 176-194.}

\title{EFFECTS OF LONG-RANGE DISPERSION IN NONLINEAR DYNAMICS
OF DNA MOLECULES}

\author{Yu.B. GAIDIDEI$^{a,b}$, S.F. MINGALEEV}

\address{$\ ^a$ Institute for
Theoretical Physics, 252143 Kiev, Ukraine.}

\author{P.L. CHRISTIANSEN, M. JOHANSSON, 
K.{\O}. RASMUSSEN}

\address{$\ ^b$ Department
of Mathematical Modelling, The Technical University of Denmark,\\
DK-2800 Lyngby, Denmark.}


\maketitle\abstracts{
A  discrete nonlinear Schr{\"o}dinger (NLS) model with 
long-range dispersive interactions  describing the dynamical structure
of DNA is proposed.  Dispersive interactions of two types: 
the power dependence $r^{-s}$ and the exponential dependence $e^{-\beta\,r}$ 
on the distance, $r$, are studied. For $s$ less than some 
critical value, $s_{cr}$, and similarly for $\beta\leq\beta_{cr}$ there 
is an interval of bistability where two stable stationary 
states : narrow, pinned states and broad, mobile states exist at each 
value of the total energy. For cubic nonlinearity the bistability 
of the solitons  occurs for dipole-dipole dispersive interaction $(s=3)$,
and for the inverse radius of the dispersive interaction $\beta\leq 
\beta_{cr}=1.67$. For increasing degree of nonlinearity, $\sigma$, 
the critical values $s_{cr}$ and $\beta_{cr}$ increase. 
The long-distance behavior of the intrinsically localized states depends 
on $s$. For $s\,>\,3$ their tails are exponential while for $2\,<\,s\,<\,3$ 
they are algebraic. A controlled switching between 
pinned and mobile states is demonstrated applying a spatially 
symmetric perturbation in the form of a parametric kick. The mechanism 
could be important for controlling energy storage and transport in DNA 
molecules.}

\section{Introduction}

Understanding the mechanisms of the functioning of biological macromolecules
(proteins, DNA, RNA, etc.) remains for now the major challenge in molecular
biology. One of the most important questions is the understanding of gene
expression. The expression of a given gene involves two steps: transcription
and translation. The transcription includes copying the linear genetic 
information into the messenger ribonucleic acid (mRNA). The information
stored in mRNA is transferred into a sequence of aminoacids using the genetic
code. mRNA is produced by the enzyme RNA-polymerase (RNAP) which binds to
the promoter segment of DNA. As a result of the interaction
between RNAP and promoter of DNA the so-called "bubble" (i.e. a state in which 
10--20 base pairs are disrupted) is formed. The disruption of
20 base pairs corresponds to investing some 100 kcal/mole (0.43 eV) 
\cite{reiss}.

In the framework of a linear model the large-amplitude motion of the bases 
was supposed to occur due to an interference mechanism \cite{loz79}. 
According to this model energetic solvent molecules kick DNA and create 
elastic waves therein. As a result of the interference of two counter 
propagating elastic waves, the base displacements may exceed the elasticity 
threshold  such that DNA undergoes a transition to a kink form which is more 
flexible. A similar approach was also proposed \cite{cm88,pr86}. The 
linear elastic waves in DNA are assumed to be strong enough to break a 
hydrogen bond and thereby facilitate the disruption of base pairs. In spite 
of the attractiveness of this theory which gives at least a qualitative 
interpretation of the experimental data \cite{geor96} 
there are following fundamental difficulties which to our opinion are inherent 
to the linear model of the DNA dynamics:
     i) The dispersive properties (the dependence of the group velocity on the 
wave-length) of the vibrational degrees of freedom in DNA will cause spreading 
of the wave packets and therefore smear the interference pattern. Furthermore, 
it has been shown \cite{hk84} that the amplitudes of the sugar and the base 
vibrations are rather large even in a crystalline phase of DNA.  Since 
the large-amplitude vibrations in the molecules and the molecular complexes 
are usually highly anharmonic their nonlinear properties can not be ignored.
     ii) Molecules and ions which exist in the solution permanently interact 
with DNA. These interactions are usually considered as white noise and their 
influence is modelled by introducing Langevin stochastic forces  into the 
equations describing the intramolecular motion. It is well known \cite{gs76} 
that stochastic forces provide relaxation of linear excitations and 
destroy their coherent properties. Equivalently the coherence length (the 
length of the concerted motions) rapidly decreases with increasing 
temperature.
     iii) DNA is a complex system which has many nearly isoenergetic 
ground states  and may therefore be considered as a fluctuating 
aperiodic system. DNA may have physical characteristics in common with 
quasi-one-dimensional disordered crystals or 
glasses. However, it is known \cite{pap71} that the transmission coefficient for a 
linear wave propagating in disordered chain decreases exponentially with the 
growth of the distance (Anderson localization). In this way it is difficult to 
explain in the framework of linear theory such a phenomenon as an action at  
distance where concerted motion initiated at one end of a 
biological molecule can be transmitted to its other end. 

The above mentioned fundamental problems can be  overcome in the framework of 
nonlinear models of DNA.
     Nonlinear interactions can give rise to very stable excitations, called 
solitons, which can travel  without being smeared out. These  
excitations are very robust and important in the coherent transfer of energy 
\cite{wa76}.  For realistic interatomic potentials the solitary waves are compressive 
and supersonic. They propagate without energy loss, and their collisions 
are almost elastic.

Nonlinear interactions between atoms in DNA can give rise to intrinsically 
localized breather-like vibration modes \cite{st88,au94}. Localized modes being 
large-amplitude vibrations of a few (2 or 3) particles, can facilitate the 
disruption of base pairs and in this way initiate conformational transitions in 
DNA. These modes can occur as a result of modulational instability of 
continuum-like nonlinear modes \cite{pou93} which is  created by energy exchange 
mechanisms between the nonlinear excitations. The latter favors the growth of 
the large excitations \cite{dp93}.

Nonlinear solitary excitations can maintain their overall shape on long 
time scales even in the presence of the thermal fluctuations. Their robust 
character under the influence of white noise was demonstrated \cite{mu90} and a 
simplified model of double-stranded DNA was proposed and explored. Quite 
recently the stability of highly localized, breather-like, excitations in 
discrete nonlinear lattices under the influence of thermal fluctuations was 
investigated \cite{ch96}. It was shown that the lifetime of a breather increases with 
increasing nonlinearity and in this way these intrinsically localized modes may 
provide an excitation energy storage even at room temperatures where the  environment 
is seriously fluctuating.

Several theoretical models have been proposed in the study of the nonlinear dynamics
and statistical mechanics of DNA (see the very comprehensive review 
\cite{grpd}). A particularly fruitful model was proposed  by 
Peyrard and Bishop \cite{pb89} and Techera, Daemen and Prohofsky
 \cite{tdp89}. In the framework of this model the DNA molecule is considered to 
consist of two chains that are transversely coupled. Each chain models one of the 
two polynucleotide strands of the DNA molecule.  A base is considered
to be rigid body connected with its opposite partner through
the hydrogen-bond potential $V(u_n)$, where $u_n$ is the 
stretching of the bond connecting the bases,  $n$, and $n=0,\pm1,\pm2,..$ is  
labelling the base-pairs. The stretching of the $n'$th base-pair is coupled with the
stretching of the $m'$th base-pair through a dispersive potential $K(u_n,u_m)$. 
The process of DNA denaturation was studied \cite{pb89,dpb93} under the assumption that 
the coupling between neighboring base-pairs is 
harmonic $K(u_n,u_{n+1})$. An entropy-driven denaturation was
investigated \cite{dpbi93} taking into account a nonlinear potential $K(u_n,u_{n+1})$
between neighboring base-pairs.  The Morse potential was chosen 
\cite{pb89,dpb93,dpbi93} as the on-site potential $V(u_n)$ but also the 
longitudinal wave propagation and the denaturation of DNA has been investigated \cite{mu90}
using the Lennard-Jones potential to describe the hydrogen bonds.

In the main part of the previous studies the dispersive interaction $K$
 was assumed to be short-ranged and a nearest-neighbor approximation was 
used. It is worth noticing, however, that one of two hydrogen bonds
which is responsible for the interbase coupling: the hydrogen bond in the
 $N-H...O$  group is characterized by a finite dipole moment.  Therefore 
a stretching of the base-pair will cause a change of the dipole moment
so that the excitation transfer in the molecule
will be due to transition dipole-dipole interaction with a $1/r^3$
dependence on the distance, $r$. It is also well known that nucleotides in
DNA are connected by hydrogen-bonded water filaments \cite{dr71,cc81}. 
In this case an effective long-range excitation transfer may occur
due to the nucleotide-water coupling. 

In the last few years the importance of the 
effect of long-range interactions (LRI) on the properties of nonlinear
excitations was demonstrated in several different areas of physics.
The effective mass of solitons in the Frenkel-Kontorova model 
with a repulsive LRI, their shapes and Peierls barriers were 
investigated \cite{bkz90}. An implicit form of solitons was obtained 
\cite{WKK93} in a sine-Gordon system with a LRI
of the Kac-Baker type \cite{Baker61,KH73} and the dependence of the soliton
width and energy on the radius of the LRI was analyzed. It was postulated \cite{VER94}
that the nonlinear term in the sine-Gordon equation has
a non-local character and novel soliton states, of topological charge zero, were
found to exist at a large enough radius of the interaction. The effects of long-range
interactions of the Kac-Baker type were studied in static and 
dynamic nonlinear Klein-Gordon models 
\cite{aekm93}, and nonlinear Schr{\"o}dinger \cite{gmcr96} continuum models. 
The effects of a long-range harmonic interaction in a chain with short-range
anharmonicity were also considered \cite{gfnm95}. It was demonstrated that the 
existence of two
velocity dependent competing length scales leads to two types of solitons with
characteristically different widths and shapes for two velocity regions
separated by a gap. 
A nonlocal NLS equation was proposed \cite{gmcr96} for systems with long-range 
dispersion effects.
In contrast to the usual NLS equation stationary solutions only
exist for a finite interval of the number of excitations. In the upper part of this
interval two different kinds of stationary solutions were found. The new one 
containing a cusp soliton was shown to be unstable. It was also pointed out that
moving solitons radiate with a wavelength proportional to the velocity. 
Quite recently \cite{ga97} we proposed a new nonlocal discrete NLS model with a 
power dependence on the distance of matrix element of dispersive interaction. It
was found that there is an interval of bistability in the NLS models with a 
long-range dispersive interaction. One of these states is a continuum-like
soliton and the other is an intrinsically localized mode. 

The goal of this contribution is to investigate the effects of long-range 
interactions on the nonlinear dynamics of the two-strand model of DNA.  
In Sec. II we present the analytical theory and the results of numerical 
simulations of stationary states of the discrete NLS model with a long-range 
dispersive interaction. We discuss the bistability phenomenon for the 
soliton solutions and their stability. In the analytical part of this section we 
use a variational approach exploiting an $\exp$-like function as a trial
function. Then, in Sec.\ III we
investigate the long-distance behavior of the nonlinear excitations and show
that intrinsically localized states of the discrete NLS model with a dispersive
interaction decaying slower than $1/r^3$ have algebraic tails. Section IV 
is devoted to the investigation of switching between bistable states. We show that a 
controlled
switching between narrow, pinned states and broad, mobile states with only small 
radiative losses
is possible when the stationary states possess an internal breathing mode.

\section{System and equations of motion}

We study  the two-strand model of DNA  which 
is described by the Lagrangian
\begin{eqnarray}
\label{1} 
L=T-K-V \; ,
\end{eqnarray}
where
\begin{equation}
T=\frac{1}{2}\,\sum_n \left(\frac{d\,u_n}{d\,t}\right)^2
\label{2}
\end{equation}
is the kinetic energy (the mass of the base-pair is chosen equal to 1),
\begin{equation} 
\label{3} 
K= \frac{1}{4} \, \sum \!\!\!\!\!\! \sum_{n, m (n \neq m)} \!\!\! 
J_{n-m}(u_m -u_n)^2 
\end{equation}
is the dispersive interbase-pair interaction of the stretchings and
\begin{equation}
\label{4}
V=\sum_n V(u_n)
\end{equation}
is the potential energy which describes an intrabase-pair interaction. 
In Eqs.\ (\ref{1})--(\ref{4}) $n$ and $m$ are site (base-pair) indices, 
$u_n$ is the base-pair stretching.  The value $u_n=0$ corresponds to the 
minimum of the intrabase-pair potential $V(u_n)$. 
We investigate the model with the following power dependence
on the distance of the matrix element of the base elastic
coupling
\begin{equation}
\label{5}
J_{n-m}=J/|n-m|^s.
\end{equation}
The constant $J$ characterizes the strength of the coupling. 
The parameter $s$ is 
introduced to cover different physical situations including the nearest-neighbor
approximation ($s=\infty$), quadrupole-quadrupole ($s=5$) and
dipole-dipole ($s=3$) interactions. 
We shall show that this equation having "tunable" properties illuminates 
both the competition between nonlinearity and dispersion and 
the interplay of long-range interactions and 
lattice discreteness. To take into account the possibility of an 
indirect coupling between base-pairs (e.g. via water filaments) we consider
also the case when the matrix element of the base elastic
coupling has the form 
\begin{equation}
\label{6}
J_{n-m}=J\,e^{-\beta|n-m|} \; ,
\end{equation} 
where $\beta$ is the inverse radius of the interaction. 

Assuming that 
\begin{eqnarray}
\label{7}
\left.
\frac{\partial^2 V(u_n)}{\partial u_n^2} \right|_{u_n=0}\,
\gg\, \left. 
\frac{\partial^j V(u_n)}{\partial u_n^j} \right|_{u_n=0}
\qquad \mbox{for} \quad j=3, 4 ...
\end{eqnarray}
i.e.\ the anharmonicity of the intrabase-pair potential is small, we will
use a rotating-wave approximation
\begin{equation}
\label{8}
u_n=\psi_n\,e^{-i\omega\,t}+c.c. \; ,
\end{equation}
where $\left. \omega=\sqrt{\frac{\partial^2 V(u_n)}{\partial u_n^2}} 
\right|_{u_n=0}$ is the frequency of the harmonic oscillations,
$\psi_n(t)$ is the complex amplitude  which is supposed to
vary slowly with time. Inserting Eq.\ (\ref{8}) into 
Eqs.\ (\ref{1})--(\ref{4}) 
and averaging with respect to the fast oscillations of the frequency $\omega$ 
we conclude that the effective Lagrangian of the system can be represented 
in the form
\begin{eqnarray}
\label{lagra}
{\cal L}=\frac{i}{2} \sum_n\left(\dot \psi_n\psi^*_n-\dot \psi_n^*\psi_n \right)-
{\cal H} \; .
\end{eqnarray}
Here $\dot{\psi_n}\equiv \frac{d }{d\tau}\,\psi_n$, $\tau=\frac{t}{2\omega}$.
\begin{eqnarray}
\label{9} 
{\cal H}={\cal K}+{\cal V}
\end{eqnarray}
is the effective Hamiltonian of the system where
\begin{equation} 
\label{10} 
{\cal K}= \frac{1}{2} \, \sum \!\!\!\!\!\! \sum_{n, m (n \neq m)} \!\!\! 
J_{n-m}|\psi_m -\psi_n|^2 
\end{equation}
is the effective dispersive energy of the excitation and
\begin{equation}
\label{11}
{\cal V}=\sum_n \left( \frac{\omega}{2\pi}\,
\int\limits_{0}^{\frac{2\pi}{\omega}}\,dt
V(\psi_n\,e^{-i\omega\,t}+c.c)-\omega^2\,|\psi_n|^2
\right)
\end{equation}
is the effective intrabase-pair potential. 
 Usually either a Morse potential
\cite{pb89,dpb93} or a Lennard-Jones potential \cite{mu90} is used to model the hydrogen 
bonds. With these potentials however it is very complicated to obtain any analytical
results. Therefore to gain insight into the problem we will use a 
simplified nonlinear potential in the form
\begin{equation}
\label{17}
{\cal V}=-\frac{1}{(\sigma +1)} \sum_n |\psi_n|^{2(\sigma +1)}
\; , 
\end{equation}
where the degree of nonlinearity $\sigma$ is a parameter which we include
to have the possibility to tune the nonlinearity as well.

From the Hamiltonian (\ref{9}) we obtain the equation of motion 
$i \dot {\psi_{n}}= \displaystyle{\frac{\partial {\cal H}}{\partial \psi_{n}^*}}$
for the wave function $\psi_{n}(\tau)$ in the form
\begin{eqnarray}
\label{12} 
i \dot{\psi_{n}}+ \sum_{m (m \neq n)} J_{n-m} (\psi_{m} - \psi_{n})+ 
|\psi_n|^{2\sigma} \psi_n=0 \; .
\end{eqnarray}
The Hamiltonian ${\cal H}$ and the number of excitations 
\begin{equation}
\label{13}
N=\sum_n |\psi_n|^2
\end{equation}
are conserved quantities. 

We are interested in stationary solutions of Eq.\ (\ref{12})
of the form 
\begin{equation}\label{sta}
\psi_n=\phi_n \exp (i \Lambda \tau)
\end{equation}
with a real shape function $\phi_n$ and frequency $\Lambda$. 
This reduces the  governing equation for $\phi_n$ to 
\begin{equation}
\label{14}
\Lambda \phi_n= J \sum_{m (m \neq n)} |n-m|^{-s} (\phi_m-\phi_n)+\phi_n^{(2\sigma 
+1)}.
\end{equation}
Thus Eq.\ (\ref{14}) is the Euler-Lagrange equation for the problem of extremizing
${\cal H}$ under the constraint $N=constant$. 

To develop a variational approach to the 
problem we use an ansatz for a localized state in the form
\begin{equation}
\label{15}
\phi_n=\sqrt{N\tanh\alpha} \exp(-\alpha |n|) \; ,
\end{equation}
where $\alpha$ is a trial parameter. The 
ansatz (\ref{15}) is chosen to satisfy automatically the normalization condition
(\ref{13}) such that the problem of extremizing ${\cal H}$ under the constraint 
$N=constant$ is reduced to the problem of satisfying the 
equation $\displaystyle{\frac{d{\cal H}}{d\alpha}=0}$.

Inserting the trial function (\ref{15}) into the Hamiltonian given by
Eqs.\ (\ref{9}), (\ref{10}), (\ref{17}) and (\ref{5}), 
and evaluating the discrete sums which enter in these equations
(see \cite{ga97} for details) we get the dispersive part of the Hamiltonian
\begin{eqnarray}
\label{19}
{\cal K}=2 N J \left\{\zeta(s)- \tanh(\alpha) F(e^{-\alpha}, s-1)
- F(e^{-\alpha}, s) \right\} \; 
\end{eqnarray}
and  the intrabase-pair potential 
\begin{eqnarray}
\label{20}
{\cal V}=-\frac{N^{\sigma+1}}{\sigma+1} \,f_{\sigma} \; ,
\quad \mbox{with} \quad 
f_{\sigma}=\tanh^{\sigma+1} [\alpha] \,\coth [(\sigma +1) \alpha ]
\; . 
\end{eqnarray}
In Eq.\ (\ref{19})
\begin{equation}
\zeta(s) = \sum_{n=1}^{\infty} n^{-s} \; 
\end{equation}
is the Riemann's zeta function and 
\begin{equation}
\label{22}
F(z, s)=\sum_{n=1}^{\infty}(z^n/n^s)
\end{equation}
is the so-called Jonqi\`{e}re's function \cite{htf53}.

\begin{figure}
\centerline{\hbox{
\psfig{figure=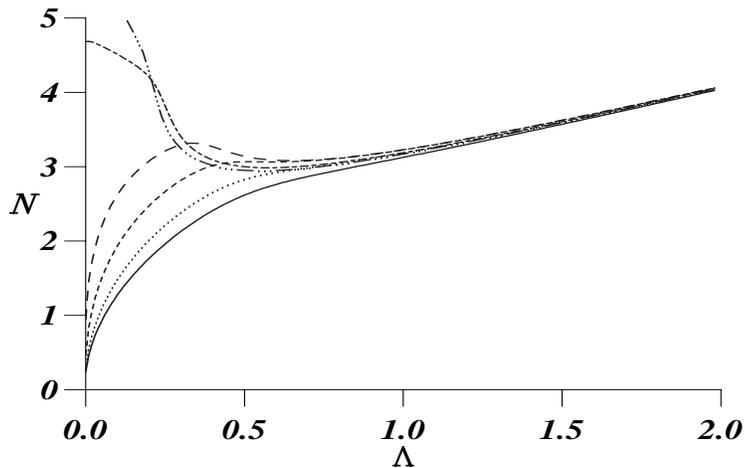,height=2.5in,width=4.7in,angle=270}}}
\caption{Number of excitations, $N$, versus frequency, $\Lambda$, 
numerically from Eq.\ (\ref{14}) for $s=\infty$ (full), 4 (dotted), 3 
(short-dashed), 2.5 (long-dashed), 2 (short-long-dashed),
1.9 (dashed-dotted).
\label{fig1}}
\end{figure}

According to the variational principle we should satisfy the condition 
$\displaystyle{\frac{d {\cal H}}{d \alpha}=0}$ which yields 
 \begin{eqnarray}
 \label{23}
 N^{\sigma}=2 (\sigma+1) J \, (\tanh (\alpha) &F(e^{-\alpha},s-2)&
  \\ \nonumber 
 + \tanh^2(\alpha) &F(e^{-\alpha},s-1)& )
 \left( \frac{d f_{\sigma}}{d\alpha} \right)^{-1} \; .
 \end{eqnarray}
As a direct consequence of Eq.\ (\ref{14}), frequency $\Lambda$ can be 
expressed as 
 \begin{eqnarray}
 \label{24}
 \Lambda=-\frac{1}{N}\, ({\cal K}+2 {\cal V})
 \end{eqnarray}
with ${\cal K}$ and ${\cal V}$ being defined by Eqs.\ (\ref{19}) and (\ref{20}).
Let us discuss first the stationary states of the system for the case 
$\sigma =1$. 

\begin{figure}
\centerline{\hbox{
\psfig{figure=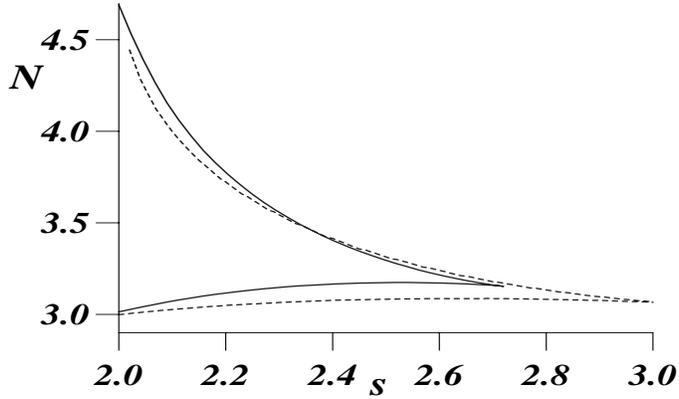,height=2.5in,width=4.7in,angle=270}}}
\caption{Shows endpoints of the bistability interval for 
$N$ versus dispersion parameter $s$. For $s=s_{cr}$ the endpoints 
coalesce. Analytical dependence (full), $s_{cr} \simeq 2.72$. 
Numerical dependence (dashed), $s_{cr} \simeq 3.03$.
\label{fig2}}
\end{figure}

Figure 1 shows the dependence $N(\Lambda)$ obtained  for $\sigma=1$  from 
direct numerical solution of 
Eq.\ (\ref{14}). A monotonic 
function 
is obtained for $s>s_{cr}$. For $s_{cr}>s>2$ the dependence becomes 
non-monotonic (of 
${\cal N}$-type) with a local maximum and a local minimum. These extrema 
coalesce at 
 $s=s_{cr} 
\simeq 3.03$.
For $s<2$ the local maximum 
disappears. The dependence $N(\Lambda)$ obtained analytically 
from Eqs.\ (\ref{23}) and (\ref{24}) is in a good qualitative agreement
with the dependence obtained numerically \cite{ga97}.
Thus the main features of all discrete NLS models with 
dispersive interaction $ J_{n-m} $ decreasing faster than $|n-m|^{-s_{cr}}$
coincide qualitatively with the features obtained in the nearest-neighbor
approximation where only
one on-site stationary state exists for any $N$. 
However, for $2<s<s_{cr}$ three stationary states with frequencies 
$\Lambda_{1}(N) < \Lambda_{2}(N) < \Lambda_{3}(N)$ there exist for 
each $N$ in the interval $[N_{l}(s), N_{u}(s)]$. In particular, this means 
that in the case of dipole-dipole interaction ($s=3$) multiple 
solutions exist.   It is noteworthy that similar results are obtained when the
dispersive interaction is in the form of the Kac-Baker potential (\ref{6}).
In this case the bistability takes place for $\beta\,\leq\,1.67$. 
 According to the theorem which was recently proven \cite{lst94},
the necessary and sufficient stability criterion for the stationary states
is 
\begin{equation}
\label{25}
\frac{dN}{d\Lambda}=\frac{d}{d \Lambda}\, \sum_n\, \phi_n^2 > 0 \; . 
\end{equation}
Therefore 
we can conclude that in the interval $[N_{l}(s), N_{u}(s)]$ there are only two
linearly stable stationary states ($\Lambda_{1}(N)$ and $\Lambda_{3}(N)$). 
The third state is unstable since $\displaystyle{\frac{d N}{d \Lambda} <0}$ 
at $\Lambda=\Lambda_2$.

\begin{figure}
\centerline{\hbox{
\psfig{figure=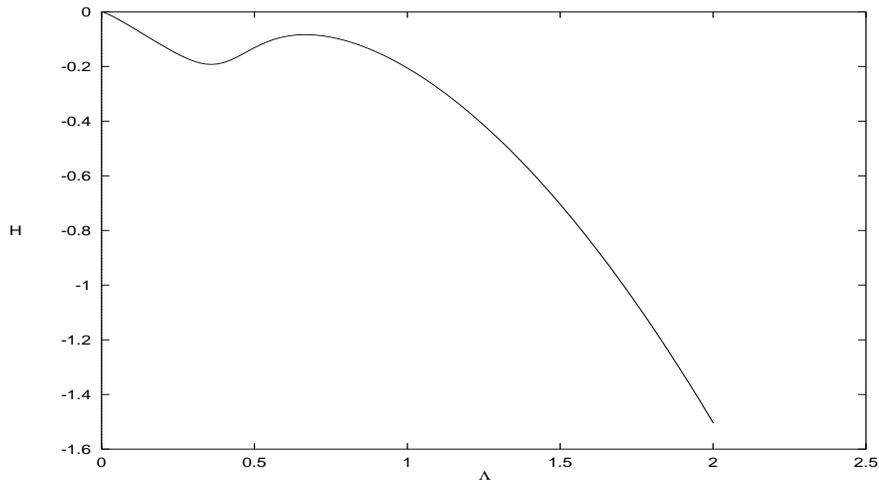,height=2.5in,width=4.7in,angle=270}}}
\caption{Hamiltonian of the system, ${\cal H}$, versus frequency, 
$\Lambda$, numerically from Eq.\ (\ref{14}) for $s=2.5$.
\label{fig3}}
\end{figure}

At the points ($\Lambda(N_l)$ and
$\Lambda(N_u)$) the stability condition is violated, since $\displaystyle{(\frac{
\partial N}{\partial \Lambda})_s}$ vanishes. Constructing the locus of the end
points we obtain the curve presented in Fig.\ 2. This curve
bounds the region of bistability. It is analogous to the critical curve in 
the van der Waals' theory of liquid-vapour phase transition \cite{ll59}. Thus 
in the present 
case we have a similar phase transition like behavior where two phases 
are the continuum states and 
the intrinsically localized states, respectively. The analog of
temperature is the parameter $s$.
Figure 3 shows the multistability phenomenon in 
terms of the Hamiltonian ${\cal H}$ of the system given 
by Eqs.\ (\ref{9})--(\ref{11}) for $\sigma=1$ and
$s=2.5$.
For $s\,<\,s_{cr}$ there is an energy interval where for each ${\cal H}$
three stationary states with different $\Lambda$ exist.
The observed bistability is very similar to the recently observed 
one \cite{lst94,Malomed}, 
where the nearest-neighbor case with an arbitrary degree of nonlinearity $\sigma$ was studied. The bistability appears 
in this case for $\sigma$ above a certain critical value.

\begin{figure}
\centerline{\hbox{
\psfig{figure=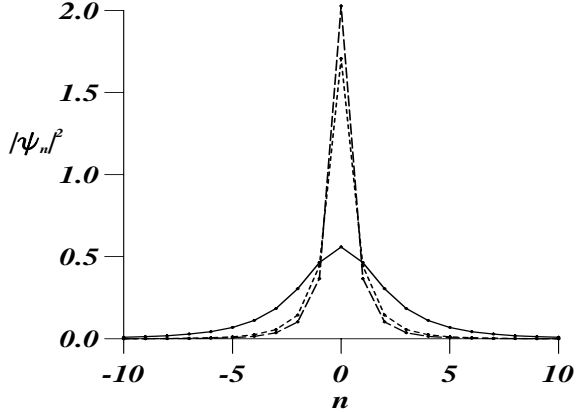,height=2.5in,width=4.7in,angle=270}}}
\caption{Shapes of the three stationary states for $s=2.5$ and $N=3.1$.
The stable: $\Lambda=0.21$ (full), $\Lambda=0.74$ (long-dashed). 
The unstable: $\Lambda=0.57$ (short-dashed).
\label{fig4}}
\end{figure}

Figure 4 shows that the shapes of these solutions differ 
significantly. The low frequency states are wide and continuum like while the
high frequency solutions represent intrinsically localized states with a width
of several lattice spacings. It can be obtained \cite{ga97} that the inverse
widths of these two stable states are 
\begin{eqnarray}
\alpha_1\, \approx \, \left(\frac{N}{8J} \right)^{1/(s-2)}= 
\left(\frac{N}{8J} \right)^{\ln \ell/(1-2\ln \ell)},~~ 
\alpha_3\, \approx \,  \ln \left(\frac{N}{J} \right) \; , 
\end{eqnarray}
where $\ell=\exp(1/s)$ is the characteristic length scale of the dispersive
interaction which is defined as a distance at which the interaction 
decreases twice. It is seen from 
these expressions that the existence of two so different soliton states for one 
value of the excitation number, N, is due to the presence of two 
different length scales in the system: the usual scale of the NLS model which 
is related to the competition between nonlinearity and dispersion 
(expressed in terms of the ratio N/J ) and the range of the dispersive 
interaction $\ell$.

Now we turn to discus stationary states of the discrete NLS model given
by Eq.\ (\ref{14}) with arbitrary degree of nonlinearity. The main properties of the
system remain unchanged, but the critical value of the dispersion parameter
$ s_{cr}$ is now a function of $\sigma$. The results of analytical consideration
confirmed by simulation  show that $s_{cr}$ increases with increasing $\sigma$. 
In particular, for $\sigma \geq 1. 4$ (the value at which discrete
symmetric ground state can be unstable in the nearest-neighbor approximation
\cite{lst94}) the bistability in the nonlinear energy spectrum occurs
even for $s\leq 6$.

\section{Tails of intrinsically localized states}

Investigating the asymptotic behavior of the excitations, it is 
convenient to rewrite Eq.\ (\ref{14}) (we consider here the case $\sigma=1$) 
in the form
\begin{equation}
\label{26}
\phi_n\,=\,\sum_m \,G_{n-m}(\Lambda) \,\phi_n^3 \; ,
\end{equation}
where
\begin{equation}
\label{27}
G_n (\Lambda)=\frac{1}{2\pi} 
\int \limits_{-\pi}^{\pi}\,d k\,\frac{\cos(k n)}{\Lambda+{\cal L}(k)} 
\end{equation} 
is the Green's function with the spectrum function
\begin{equation}
{\cal L}(k) = 2 \sum_{n=1}^{\infty} J_n (1-\cos(kn)) \; .
\end{equation} 
Deriving the asymptotic expressions for the Green's 
function (\ref{27}) \cite{ga97}  we obtain
that the tails of the intrinsically localized states are given by the expressions
\begin{equation}
\label{28}
\phi_n \rightarrow  
 \sqrt{\frac{(\Lambda+1)^3 \zeta(s)} {2 \Lambda \zeta(s-2)}} 
\exp\left (-\sqrt{\frac{2 \Lambda \zeta(s)}{\zeta(s-2)}} |n| \right)\; , \quad  
s > 3 \; ,
\end{equation}
 \begin{equation}
 \label{29}
\phi_n \rightarrow \frac{(\Lambda+1)^{\frac{3}{2}}}{ \Lambda^2 }|n|^{-s} \; , 
\quad  
2 < s < 3 \; ,
\end{equation}
 for $|n|\rightarrow \infty$. Thus we can conclude here that only in the case
 of the short-range dispersion ($s\,>\,3$) the tails of intrinsically localized
 states have a usual exponential form. In the systems with long-range 
 dispersive interactions these states have algebraic tails. Figure 5 shows 
 the long-distance behavior of the intrinsically localized states for $s=2.5$ and 
 different values of the frequency $\Lambda$. It is seen that the form of the tails 
 predicted by Eq.\ (\ref{29}) is in a good agreement with the results of 
 numerical simulations.
 
\begin{figure}
\centerline{\hbox{
\psfig{figure=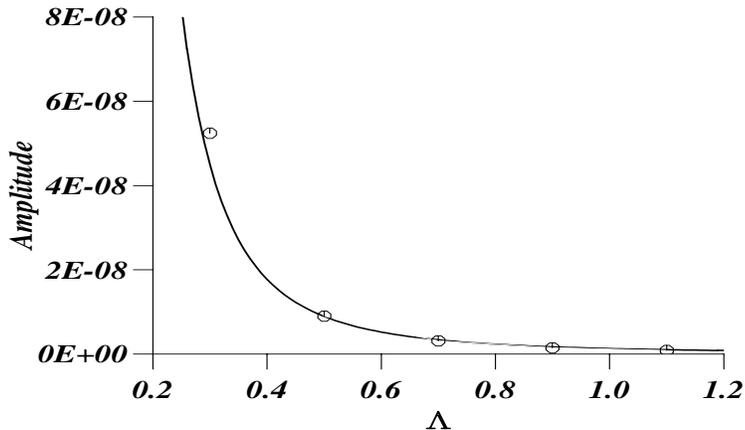,height=2.5in,width=4.7in,angle=270}}}
\caption{Amplitude in the tail of the stationary state for $s=2.5$ 
and $n=450$. Numerical (circles) and Eq.\ (\ref{29}) (full).
\label{fig5}}
\end{figure}

The long-distance behavior of the intrinsically localized states 
may play an essential role in the thermodynamic properties of the DNA
 molecule because in systems
where the interaction decays algebraically it can be responsible for the
appearance of new thermodynamically stable states (for
example, quite recently \cite{pbp97}
the existence of the Neel order in the ground state
of Heisenberg antiferromagnetic chains with algebraic long-range interactions
was proven).

\section{Switching between bistable states}

Having established the existence of bistable stationary states in the 
nonlocal discrete NLS system, a natural question that arises concerns
 the role of these states in the full dynamics of the model. In particular,
it is of interest to investigate the possibility of switching between
 the stable states under the influence of external perturbations, and what
type of perturbations could be used to control the switching.
Switching of this type is important in the description of nonlinear transport
and storage of energy in biomolecules like the DNA, since a mobile 
continuum-like excitation can provide action at distance while the switching
to a discrete, pinned state can facilitate the structural changes of the DNA
\cite{grpd,geor96}. As it was shown recently \cite{mj97} the switching will
occur if the system is perturbed in a way so that an internal, spatially 
localized and symmetric mode ('breathing mode') of the stationary state is
excited above a threshold value.

To investigate the time evolution of an initially small perturbation
$\epsilon_n(0)$ of the stationary state (\ref{sta}) we write
\begin{equation}
\label{s1}
\psi_n(\tau)=(\phi_n+\epsilon_n(\tau))\,e^{i\Lambda\, \tau}
\end{equation}
Decomposing $\epsilon_n(\tau)$ into real and imaginary parts, $\epsilon^{(r)}_n$ and
$\epsilon^{(i)}_n$, we obtain from Eq.\ (\ref{12}) with $\sigma=1$ in the
linear approximation
\begin{equation}
\frac{d}{d\tau}\,\left(\begin{array}{c}\epsilon^{(r)}_n\\ 
\epsilon^{(i)}_n\end{array}\right)= {\cal M}\left(\begin{array}{c}\epsilon^{(r)}_n\\ 
\epsilon^{(i)}_n\end{array}\right)
\equiv\,\left(\begin{array}{clcr}
 0&H^+\\ \label{s2}
 -H^-&0
 \end{array}\right)\left(\begin{array}{c}\epsilon^{(r)}_n\\ 
\epsilon^{(i)}_n\end{array}\right)
\end{equation}
where, for a system with $M$ sites, $H^+$ and $H^-$ are $M\times M$ matrices defined by
\begin{equation}
\label{s3}
H^{\pm}_{ij}=\left(\Lambda-(2\mp\,1)\phi_i^2+2\right)\,\delta_{i,j}-J_{i-j},
\end{equation}
with $J_0=0$. By definition the stationary solution is linearly stable if the 
perturbation
$\epsilon_n(\tau)$ as calculated from Eq.\ (\ref{s2}) remains bounded for all times.
Linear stability is then equivalent to the matrix ${\cal M}$ having no
eigenvalues with a positive real part. Changing some parameter (e.g.
$\Lambda$), a stable state might become unstable. The 'direction' in which
an initial perturbation will grow is then determined by the eigenvector
corresponding to the eigenvalue of ${\cal M}$ with a positive real part.
We will in sequel mainly discuss the case when the matrix element of 
base elastic coupling $J_{n-m}$ decreases exponentially  with the distance
$|n-m|$ (see Eq.\ (\ref{6})) with the inverse radius of the interaction 
$\beta=1$. For such value of $\beta$ the multistability occurs in the interval
$3.23\,\leq\,N\,\leq \,3.78$. It is worth noticing, however, that the
scenario of switching described below remains qualitatively unchanged
for all values of $\beta\,\leq\,1.67$, and also for the algebraically
decaying dispersive coupling with $2\,\leq\,s\,\leq\,3.03$.

The study \cite{mj97} of the eigenvalue problem for the matrix 
${\cal M}$ showed the existence of a 
spatially symmetric internal breathing mode for both the narrow and broad 
components of the bistable state. Furthermore, the low frequency (broad)
component also possesses a spatially antisymmetric translational ("pinning") 
mode \cite{ding}.
Since the appearance of a translational mode implies that the stationary state
gains mobility \cite{ding}, the continuum-like state will have a high
mobility.

\begin{figure}
\centerline{\hbox{
\psfig{figure=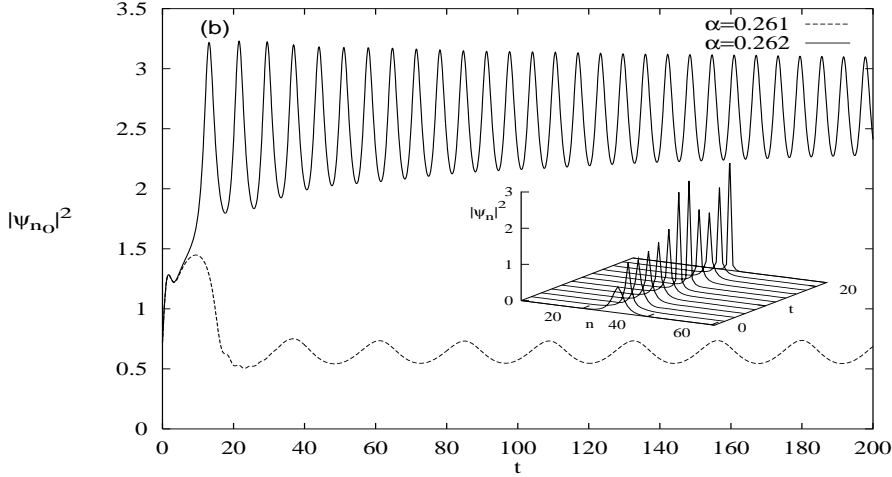,height=2.5in,width=4.7in,angle=270}}}
\caption{Switching from continuum-like to discrete state for $\beta=1$.
The initial state $\phi_n$ has the frequency $\Lambda\simeq 0.31$ and
$N=3.6$. The time evolution of $|\psi_{n_0}(\tau)|^2$ when a phase torsion 
is applied to the center site with $\theta=0.261$ (lower curve) and 
$\theta=0.262$ (upper curve), respectively; inset shows time 
evolution of $|\psi_{n}(\tau)|^2$ for $\theta=0.262$.
\label{fig6}}
\end{figure}

An illustration of how the presence of an internal breathing mode can affect 
the dynamics
of a slightly perturbed stable stationary state is given in Fig.\ 6. To excite
the breathing mode we apply a spatially symmetric, localized perturbation,
which we choose to conserve the number of excitations in order not to change 
the effective 
nonlinearity of the system. The simplest choice, which we have used in the 
simulations shown here, is to kick the central site $n_0$ of the system at
$t=0$ by adding a parametric force term of the form
$\theta\delta_{n,n_0}\delta(\tau)\psi_n(\tau)$ to the left-hand-side of
Eq.\ (\ref{12}). A possible physical motivation of the appearance of such 
kind of parametric kick may be 
the following. It is well known that  biomolecules in cells interact with  
solvent molecules and ions (ligands).  At the position where the ligand
links to a biomolecule the potential energy of the biomolecule
changes. This manifests itself in a local change of the vibration frequency
of biomolecular units. In our case
it means that the frequency $\omega$ of the internal base-pairs 
oscillations locally changes and the potential energy (\ref{4}) should be
replaced by
\begin{equation}
\label{s4}
V=\sum_n\left( V(u_n)+\frac{1}{2}\delta_{n n_0}\theta(\tau)\,u^2_n\right)
\end{equation}
where $n_0$  is the site where the ligand interacts with the biomolecule,
$\theta$ is a constant which characterizes the energy interaction between
the ligand and biomolecule. It may be a function of time because the ligand
may attach or detach at different time moments. If we use the rotating wave
approximation (\ref{8}) the NLS model given by Eq.\ (\ref{lagra})
with the additional term in Eq.\ (\ref{11}) being 
\begin{equation}
{\cal V}_{int}=\sum_n\, \delta_{n\,n_0}\theta(\tau)|\psi_n(\tau)|^2.
\end{equation}
Assuming that
\begin{equation}
\theta(\tau)=\sum_j\,\theta_j\,\delta(\tau-\tau_j)
\end{equation}
we obtain the model where the attachment and detachment of ligands are
considered as kicks which occur at the time moments $\tau_j$.

As can be easily shown, this perturbation affects only the site
$n_0$ at $\tau=0$, and results in a 'twist' of the stationary state at this state
with an angle $\theta$, i.e. $\psi_{n_0}(0)=\phi_{n_0}\,e^{i\theta}$. The
immediate consequence of this kick is, as can been deduced from the form
of Eq.\ (\ref{12}), that $\frac{d}{d\tau}\left(|\psi_{n_0}|^2\right)$ will be
positive (negative) when $\theta\,>\,0$ ( $\theta\,<\,0$). Thus, to obtain
switching from the continuum-like state to the discrete state we choose 
$\theta\,>\,0$, while we choose  $\theta\,<\,0$ when investigating switching
in the opposite direction. We find that in a large part of the multistability
regime there is a well-defined threshold value $\theta_{th}$, 
such that when the initial phase torsion is smaller than $\theta_{th}$, 
periodic, slowly decaying 'breather' oscillations around the initial state
will occur, while for strong enough kicks (phase torsions larger than
$\theta_{th}$) the state switches into the other stable stationary state.

It is worth remarking that the particular choice of perturbation is not
important for the qualitative features of the switching, as long as  there is
a substantial overlap between the perturbation and the internal breathing
mode. We believe also that the mechanism for switching described here
can be applied for any multistable system where the instability is connected with
a breathing mode. For example, we observed \cite{mjo97} 
a similar switching behavior in
the nearest neighbor discrete NLS equation with a higher degree of
nonlinearity $\sigma$, which is known \cite{lst94} to exhibit multistability.

\section{Conclusion}

We have proposed a new nonlocal discrete nonlinear Schr{\"o}dinger
model for the dynamical structure of DNA with long-range ($r^{-s}$ and
$e^{-\beta r}$) dispersive interaction. We have shown that there is
a multistability in the spectrum of stationary states of the model with
a long-range  dispersive interaction $s<s_{cr}$ ($\beta<\beta_{cr}$).
There is an energy interval where
two stable stationary states exist at each value of the Hamiltonian
 ${\cal H}$. One of these
states is a continuum-like soliton and the other one is an intrinsically
localized mode. The existence of the bistability phenomenon in the NLS models
with a nonlocal dispersion is a result of the competition of two length scales
which exist in the system: the scale related to the competition between
nonlinearity and dispersion, and the scale related to the dispersion interaction.

We found that the critical value of the dispersion parameter $s_{cr}$
for the on-site stationary state in the case of cubic nonlinearity
exceeds $3$. This means that the bistable behavior may occur in the case of
DNA  where the stretching motion of base-pairs is accompanied by a change
of their dipole moments.

We have shown that the long-distance behavior of intrinsically localized
states in discrete NLS models with a nonlocal dispersion
depends drastically on the value of the dispersive parameter $s$. 
Only for short-range dispersions the excitation wave functions decay
exponentially. In the systems where the matrix element of
base elastic coupling 
depends on the distance slower than $1/r^3$ the nonlinear excitations have
algebraic tails. The long-distance behavior may be important for
the thermodynamics of DNA since it provides long-range order in
one-dimensional systems.

We have shown that a controlled switching between narrow, pinned states
and broad, mobile states is possible. Applying a perturbation in the form 
of parametric kick, we showed that switching occurs beyond some well-defined
threshold value of the kick strength.

The particular choice of perturbation is not
important for the qualitative features of the switching, as long as  there is
a substantial overlap between the perturbation and the internal breathing
mode. Thus, we believe that the mechanism for switching described here
can be applied for any multistable system where the instability is 
connected with a breathing mode.
The switching  phenomenon could be important for controlling energy
storage and transport in DNA molecules.

\section*{Acknowledgments}

 Yu.G. and S.M.  acknowledge support from the Ukrainian Fundamental 
Research Foundation (grant \#2.4 / 355). Yu.G.
acknowledges also partial financial support from SRC QM
"Vidhuk". M.J. acknowledges financial
support from the Swedish Foundation STINT.


\section*{References}
\begin{thebibliography}{99}

\bibitem{reiss} C. Reiss, in {\sl Nonlinear Excitations in Biomolecules},
Ed.: M. Peyrard (Springer-Verlag Berlin, Heidelberg, Les Editions de
Physique Les Ulis, 1995), 29.
\bibitem{loz79} Lozansky {\em et al}, in {\sl Stereodynamics of Molecular 
Systems}, Ed.: R.H. Sarma (Pergamon Press, 1979), 265.
\bibitem{cm88} K.C. Chou and B. Mao, Biopolymers {\bf 27}, 1795 (1988).
\bibitem{pr86} E.W. Prohovsky, in {\sl Biomolecular Stereodynamics IV}, 
Eds.: R.H. Sarma and M.H. Sarma (Adenine Guilderland N.Y., 1986), 21.
\bibitem{geor96} Georghiou {\em et al},  Biophysical J. {\bf 70}, 
1909 (1996).
\bibitem{hk84} S.R. Holbrook and  S.H. Kim,  J. Mol. Biol. {\bf 173}, 
361 (1984).
\bibitem{gs76} Yu.B. Gaididei and A.A. Serikov, Theor. and  Math. Phys. 
{\bf 27}, 457 (1976).
\bibitem{pap71} G.C. Papanicolaou, J. Appl. Math. {\bf 21}, 13 (1971).
\bibitem{wa76} M. Wadati, J. Phys. Soc. Jpn. {\bf 38}, 673 (1976).
\bibitem{st88} A.J. Sievers and S. Takeno, Phys. Rev. Let. {\bf 61}, 
970 (1988).
\bibitem{au94}  R.S. MacKay and S. Aubry, Nonlinearity {\bf 7}, 1623 (1994).
\bibitem{pou93} J. Pouget {\em et al},  Phys. Rev. B {\bf 47},14866 (1993).
\bibitem{dp93} T. Dauxois and M. Peyrard, Phys. Rev. Let. {\bf 70}, 
3935 (1993).
\bibitem{mu90} V. Muto {\em et  al},  Phys. Rev. A {\bf 42}, 7452 (1990).
\bibitem{ch96} P.L. Christiansen {\em et al}, Phys. Rev. B {\bf 55}, 
5729 (1997).
\bibitem{grpd} G. Gaeta {\em et al}, Riv. N. Chim. {\bf 17}, 1 (1994).
\bibitem{pb89} M. Peyrard and A.R. Bishop, Phys. Rev. Lett. {\bf 62}, 
2755 (1989).
\bibitem{tdp89} M. Techera, L.L. Daemen, and E.W. Prohofsky, Phy. Rev. A
{\bf 40}, 6636 (1989).
\bibitem{dpb93}  T. Dauxois, M. Peyrard, and  A.R. Bishop, Phys. Rev. E 
{\bf 47}, 684 (1993).
\bibitem{dpbi93} T. Dauxois, M. Peyrard, and  A.R. Bishop, Phys. Rev. E 
{\bf 47}, R44 (1993).
\bibitem{dr71} U. Dahlborg and A. Rupprecht, Biopolymers {\bf10}, 849 (1971).
\bibitem{cc81} G. Corongiu and E. Clementi, Biopolymers {\bf 20}, 551, (1981).
\bibitem{bkz90} O.M. Braun, Yu.S. Kivshar, and I.I. Zelenskaya, 
Phys. Rev. B {\bf 41}, 7118 (1990).
\bibitem{WKK93} P. Woafo, J.R. Kenne, and T.C. Kofane, 
J. Phys. Condens. Matter {\bf 5}, L123 (1993).
\bibitem{Baker61}
G.A. Baker Jr, Phys. Rev. {\bf 122}, 1477 (1961).
\bibitem{KH73}
A.M. Kac and B.C. Helfand, J. Math. Phys. {\bf 4}, 1078 (1972).
\bibitem{VER94}
L. Vazquez, W.A.B. Evans and G. Rickayzen, Phys. Lett. A
{\bf 189}, 454 (1994).
\bibitem{aekm93}  G.L. Alfimov {\em et al},  Chaos {\bf 3}, 405 (1993).
\bibitem{gmcr96} Yu.B. Gaididei {\em et al}, Phys.Lett. A {\bf 222}, 
152 (1996);
Yu.B. Gaididei {\em et al.}, Phys. Scr. {\bf T67}, 151 (1996). 
\bibitem{gfnm95}  Yu.B. Gaididei {\em et al}, Phys. Rev. Lett. {\bf 75}, 
2240 (1995).
\bibitem{ga97} Yu.B. Gaididei {\em et al}, Phys. Rev. E {\bf 55}, 
6141 (1997).
\bibitem{htf53} W. Magnus, F. Oberhettinger and R.P. Soni, 
{\sl Formulas and Theorems for the Special Functions of Mathematical Physics} 
(Springer-Verlag, Berlin, 1966).
\bibitem{lst94} E.W. Laedke, K.H. Spatschek, and S.K. Turitsyn, 
Phys. Rev. Lett. {\bf 73}, 1055 (1994).
\bibitem{ll59} L.D. Landau and E.M. Lifshitz, {\sl Statistical Physics} 
(Pergamon Press, London, 1959).
\bibitem{Malomed} B. Malomed and M.I. Weinstein, Phys. Lett. A {\bf 220}, 
91 (1996). 
\bibitem{pbp97} J.R. Pareira, O. Bolina, and J.F. Perez, J.Phys. A {\bf 30},
1095 (1997).
\bibitem{mj97} M. Johansson, Yu.B. Gaididei, P.L. Christiansen, 
and K.{\O}. Rasmussen, Phys. Rev. E {\bf 57}, 4739 (1998).
\bibitem{ding} Ding Chen, S. Aubry, and G. Tsironis, Phys. Rev. Lett. 
{\bf77}, 4776 (1996).
\bibitem{mjo97} M. Johansson {\em et al}, Physica D {\bf 119}, 115 (1998).

\end{thebibliography}
\end{document}